\documentclass[aps,showpacs,preprintnumbers,amsmath,amssymb, 
twocolumn, tightenlines]{revtex4}

\usepackage[dvips]{graphicx}\usepackage{bm}
\usepackage{amsmath}
\usepackage{dcolumn}
\usepackage{bm} 
\newcommand{\be}{\begin{eqnarray}}
\newcommand{\ee}{\end{eqnarray}}
\def\N{${\cal N}\,\,$}

\begin{document}
\title{A ``Domain Wall'' Scenario for the AdS/QCD  }

\author{ E.Shuryak}
\affiliation{ 
Department of Physics and Astronomy,  Stony Brook University , 
Stony Brook NY 11794-3800, USA
}

\date{\today}
\begin{abstract}
We propose a scenario for bottom-up gravity dual picture of QCD-like
theories, which consists of two near-AdS$_5$ domains separated by the 
``domain wall'' at which the effective coupling relatively rapidly switches
from small perturbative value at its UV side to strong   at
its IR side. Its consequence are corresponding jumps in 5-d masses
of the bulk fields, related to different anomalous dimensions of the corresponding operators. Inclusion of both weakly and strongly coupled parts of bulk
wave functions allows for consistent inclusion of QCD hard processes.
 We further discuss how transitions from strong to weak coupling in hard
 observables should look like, exemplified by
 pion diffractive dissociation,  which  seems to show this transition experimentally.
Then we propose a dynamical mechanism for this jump  related to instantons,
which are point-like bulk objects located at/near the domain wall.
We further argue that in the limit of
 large number of
colors $N_c\rightarrow\infty$ 
the ``domain wall" is becoming a true singularity.
Instanton-induced contributions to correlators  and 
partition function of the instanton ensemble is reformulated in the $AdS_5$
language.
Among other applications  are lowest mesons as collective
vibrations of the ``domain wall''. 
\end{abstract}
\maketitle
\vspace{0.1in}
\section{Introduction}
\subsection{The bottom-up approach to gravity dual}

  Discovery of the AdS/CFT correspondence \cite{Maldacena:1997re} between $\cal N$=4
SYM in 4 dimensions and string theory in $AdS_5*S_5$ geometry
had finally fulfilled the long-standing promise of exact 
gauge-string correspondence. However,  $\cal N$=4 SYM theory
is very far from QCD: it is non-confining and conformal.

Furthermore, the simplest and most useful form of the
correspondence takes place in the large number of colors
$N_c\rightarrow \infty$ limit, in which t'Hooft gauge coupling
$\lambda=g^2N_c$ is large. One set  of AdS/CFT applications
 is related to the finite-temperature QCD and
recent discovery of Quark-Gluon Plasma at RHIC, apparently
being  strongly coupled
 \cite{liquid}. 
In this phase both QCD confinement and running coupling are of secondary
importance, while AdS/CFT results are in  agreement
with phenomenology, see e.g. recent review\cite{Shuryak:2007qs}. 

 One way to develop further those fascinating tools is to
look for a correspondence including
 theories with running coupling, e.g.
cascading theories
\cite{Herzog:2004tr}.
Another -- known as {\em top-down approach} -- is to invent 
brane constructions reproducing  QCD-like theory
 as an effective theory in infrared. Important issues
are implementation of flavor and chiral symmetry breaking
in this context \cite{chiral_ads}: currently the most widely used example of it is the
  Sakai-Sugimoto model \cite{Sakai:2004cn}.

This paper however is not about QGP or top-down models
but about an alternative approach  -- known
now as {\em bottom-up} or {\em AdS/QCD}  --
 building a  model based on known features of
the QCD vacuum and hadronic physics, formulated using  a holographic 5-d
language.

  From the onset of this paper we would like to emphasize the difference
  between  (i) a holographic approach in general and  (ii) classical gravity
  dual description in particular. The former (i) uses the 5-th coordinate as a
  general (energy) scale variable $z\sim 1/\mu$ in a sense in which it appears
  in the renormalization group, with $\mu$ being a normalization scale. 
  A notion of a hadronic wave function $\psi(z)$  is introduced, as  the
  probability amplitude for a hadron to have certain physical size.
  The  more specific classical (super)gravity description -- based on gauge-string duality
  and coupling inversion -- may or may not be applicable at certain $z$, depending on whether the
  gauge coupling is or is not large.
   AdS/QCD
     assumes that gauge theory is strongly coupled in some region of
     scales .  Even if so, some part of the
      $z$ domain corresponds to weak coupling, and  
     the problem to be discussed below
      is how to unite both in the same framework.
  Solving this problem is key for such important applications as e.g.
hard  exclusive processes, for which existence of a single  wave function $\psi(z)$
in the whole space, with single  well-defined normalization procedure,
is obviously required. 

Let us now briefly review the papers which established AdS/QCD,
pointing out where our approach would differ with what was
used there.

   {\bf Modifications at large $z$ -- Infra Red }(IR)   were introduced
 to include the effect of confinement, present in QCD and absent in AdS/CFT.
The first crude way to do so
  (introduced by Polchinski and Strassler) was done via
the IR  cutoff of the $AdS_5$
space above some $z>z_{IR}$. If this is done,
the bulk fields propagating in the bulk
get  quantized, which
generates a 4-d hadronic spectrum . It appears that this spectroscopy is not
 bad for a first-order model. One widely known good feature is
 Regge behavior of particle masses at large spins, see e.g.
\cite{Brodsky:2007pt}. The model can naturally be appended by
quark masses/condensates , see \cite{Erlich:2005qh}.
 Good news about this simple model
 are not only related with hadronic masses: sizes of the states
 and related coupling constants to point operators are in some examples
very nice as well. Very recently 
(after the first version of the present paper was already posted)
 Schafer \cite{Schaefer:2007qy} have demonstrated 
 that the whole Euclidean
 vector and axial correlators  $V(\tau),A(\tau)$ 
 built out of these modes  reproduce experimental data within 10-20\% 
accuracy, at any Euclidean time $\tau$. 
(These results are extremely similar
to what was obtained by Schafer and myself in much more sophisticated
 instanton model \cite{Schafer:2000rv}. The remaining
 10\% deviations, at least at small distances,
are clearly just neglected
 perturbative correction $1+\alpha_s/\pi+...$.)
We will however argue below that vector/axial channels are
in fact $ exceptions $ rather than the rule,
while generic correlators should
not be well described by  this simplest model.

 Further improvements of Regge phenomenology to radial excitations 
  motivated
Karch et al \cite{Karch:2006pv} (below KKSS) to introduce the
so called ``soft wall'' with a  dilaton potential quadratic in $z$.
Independent argument for quadratic wall was given in my paper
\protect\cite{Shuryak:2006yx} to which we return below.
This soft wall was interpreted  as a pure
 geometrical factor $\sim exp(C z^2)$ in metric
\cite{Andreev:2006eh}.
 Although the applications discussed in these way are
all interesting, it is hard to agree with
pure geometric 5-d modeling of confinement.
In particular, such metric  
 acts in the same way on
electric and magnetic strings, providing confinement to both.
 The dilaton field
naturally  has different coupling to electric and magnetic
(D1) strings, which allows only the former to be confined
and the latter screened, see  
``improved AdS/QCD'' by Gursoy, Kiritsis and Nitti (GKN)
\cite{Gursoy:2007er,Gursoy:2007cb}.

{\bf Modifications at small $z$ -- Ultra Violet (UV)} are also   necessary since
here QCD is weakly coupled.
Evans et al \cite{UV_cutoff} proposed to introduce an
 UV cutoff of the AdS space, simply removing some part of it
 $z<z_{UV}$ and reformulating the boundary conditions. Although that  improves
masses of the vector mesons, reconciling the lightest n=0 $\rho$
with a different  trend for  radial excitations $n=1..4$,
it goes with a heavy price. 
As we will argue below in detail,
simply cutting off the weakly coupled domain is not an
acceptable option if one hope to address
hard scattering. 
 
 In fact cutting off the space
in IR, $z>z_{IR}$  is  not acceptable as well.
It may be enough for crude features like Regge trajectories, 
but would fail to address other important issues. I particular, 
 QCD practitioners  spent a lot time trying to understand why 
 is it that quark masses changing
from say $m_q\sim m_s\sim  100\, MeV$ to $m_u,m_d\sim few MeV$
induce nontrivial changes in many observables, such
as Dirac eigenvalues,
nucleon
 structure functions, magnetic moments. In the gravity
dual language it means that one has  to accommodate
 those low mass scales, or  $z$
of the order of  several fm at least: we will
return to this in section \ref{sec_inst_ads}.
In the instanton context
 the weak coupling  is obviously a dilaton-based potential 
related to the beta function: such potential
was included in \cite{Shuryak:2006yx}, and we return to this below.
GKN \cite{Gursoy:2007er,Gursoy:2007cb} have proposed a systematic
approach 
toward building a gravity dual model to QCD,
including the metric, dilaton and axion  in the common
Lagrangian.  A clever tric allows to incorporate
the beta function (which demands first order differential eqn)
with Lagrangian formalism which demands second order ones.
Solution of the resulting equations of motion
are classified according to different dependence of these bulk fields
 on the 5-th holographic $z$
coordinate. A new issue raised by GKN
 (to be discussed in section \ref{seq_open}) 
is z-dependent axion field.

\begin{figure}[t]
 \includegraphics[width=8cm]{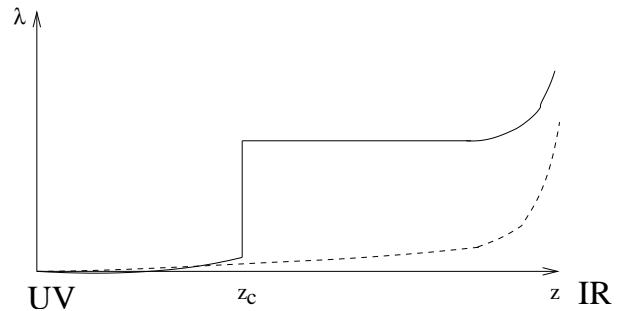}
  \caption{Schematic dependence of the t'Hooft coupling $\lambda$
on the holographic 5-th coordinate $z$. 
  } \label{fig_jump}
\end{figure}

The main subject
of the present paper is {\bf how coupling runs at 
intermediate $z$}. While
 GKN simply suggested a smooth interpolation between
IR and UV  (the dashed
line in Fig.\ref{fig_jump}), we 
expect 
a very rapid growth -- a jump --  from weak to strong coupling,
 near some particular $z$ we call the {\em domain wall}  $z_{dw}$, 
 schematically shown by the solid line
in Fig.\ref{fig_jump}. 
We further suggest as an approximation 
two ``domains'', the weakly and strongly coupled ones,
separated by this ``domain wall''. 
While we will argue that the wall is rather narrow
already for real QCD with $N_c=3$ colors,  it should become a sharp jump
in the large-$N_c$ limit.

\subsection{Old phenomenology issues:
 the ``chiral scale'' versus $\Lambda_{QCD}$}

Let me start  by reminding some 30-year-old puzzles, containing 
the essence of the issues to be discussed. Basic chiral dynamics -- 
 chiral symmetry breaking by  the quark-antiquark condensate, pions as 
Goldstone modes,  their effective Lagrangian
--   were developed in 1960's. Nambu-Iona-Lasinio model
of 1961  introduced the ``chiral scale''
\be \Lambda_\chi=4\pi f_\pi\approx 1.2 \, GeV  \label{eqn_chiral}
\ee as a boundary below which
(at $Q<\Lambda_\chi$) the hypothetical
4-fermion interaction had to exist.

With the advent of QCD  in 1970's,  observation of
scaling violation and other effects the details of  the running
coupling were clarified. It turns out that 
 that $\Lambda_{QCD}\sim 1 \, fm^{-1}\sim 200 MeV$ is quite small.
(Precise definition depends on
  subtraction scheme, but basically it is the place of
the Landau pole where the QCD coupling blows up
if one would extrapolate the one-loop pQCD running to IR.)
That lead to a puzzle: {\em why
don't these two scales  match?} 
 In particular, at $Q\sim \Lambda_\chi$ the perturbative
coupling is still too small 
to account for the necessary
strength of the NJL-type 4-fermion 
interaction  to achieve  chiral symmetry breaking. 
 Thus, already 30 years ago it was clear that some important
physics at the scale $Q\sim \Lambda_\chi$ was missing.

    The QCD sum rules  \cite{SVZ} have shown
that non-perturbative splittings of the correlators for
 some channels (such as vector and axial ones) can be
described by
 the operator product expansion (OPE). 
However, 
  much stronger non-perturbative effects
were found  for  all spin zero
channels \cite{Novikov:1981xj}. The largest are
 in  gluonic spin zero
channels, where the perturbative boundary scale is located as high as 
 $Q^2>\Lambda_{glue}^2\sim 20\, GeV^2$.
These puzzles were resolved  by realization \cite{Shuryak:1981ff}
that small-size  instantons $\rho\sim 1/3 \, fm$   are
very abundant in the QCD vacuum. Some brief overview will be given
in section \ref{sec_inst}, for review see see \cite{Schafer:1996wv}.

\subsection{Running coupling,  from lattice
and phenomenology}

The usual plots one finds in textbooks follow the coupling
down  to
$Q=2\, GeV$, where $\alpha_s(Q)\approx 1/3$. Extrapolation of these
curves down to $Q=1\, GeV$ can perhaps be still trusted, but
what is beyond this scale remains
unclear and  depends on the definition used.

\begin{figure}[htb]
\includegraphics{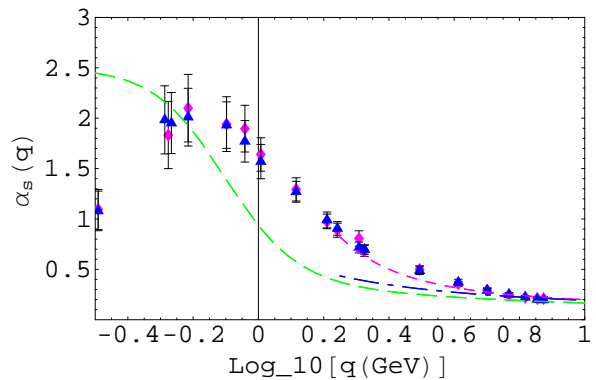}
\caption[]{The running coupling constant
from the 2-flavor unquenched data of MILC$_c$
collaboration, with lattice spacing $a=0.12$fm, $\beta_{imp}=6.76$(triangles) and 6.83(diamonds).  }
\label{fig_furui_nakajima}
\end{figure}

Having said that, let us still have a look at some results going well beyond
this scale, looking for qualitative hints.
One example, taken from \cite{Furui:2005bu}, is
 shown in Fig.\ref{fig_furui_nakajima}. It follows the
gluon-ghost-ghost vertex function in Landau gauge, for 2-flavor
QCD. The curves are different extrapolations based on pQCD.
One can see that lattice data (points) rather rapidly depart from it
already at $q\sim 2 \, GeV$ upward. Quite remarkable
feature of these results is that the coupling stays flat in some
 ``conformal window'' $q=.3-1 GeV$.
 Note that
the coupling becomes in IR rather strong indeed, 
especially if translated into the units of t'Hooft coupling
\be \lambda_{q<1\, GeV}=g^2N_c\approx 80 \ee

Another issue important for what follows is how the coupling
 runs for gauge theories with increasing number of colors $N_c$. 
 On general grounds one expects running to become more rapid: 
  the question is whether some discontinuities
 may appear at finite or infinite $N_c$.
 In recent paper by Allton et al \cite{Allton:2007py} it was found that  
 there exist a lattice bulk weak-strong phase transition starting from $N_c=5$,
 well seen in the $N_c=8$ example mainly discussed in this work. What is worth
 mentioning is that the lattice spacing corresponding to unstable (mixed phase)
 situation between weak and strong coupling is in the 
 range :
 \be  a \approx {(.4..1) \over \sqrt{\sigma}} \approx 0.2..0.5 fm  \ee 
It is precisely the range where we propose to place a weak-strong domain wall.
It is also the range in which the instanton sizes are located in this
theory. (Below we will speculate that rapid weak-strong transition
at corresponding scale is related to instantons.)
 

\section{Spectroscopy and hard processes in the two-domain picture}
\label{sec_hard}
\subsection{ From bulk to boundary spectroscopy}

General rules of ``holography'' connect
operators on the boundary  to fields $\phi(z)$
propagating in the bulk. The 
 conformal dimensions of   boundary operators 
are related to 5-d
 masses of bulk fields. 
We will denote those $M_5$ to be distinguished from
the usual 4-d masses denoted by $m_4$. The operator dimension $\Delta $
matches the wave functions asymptotics at small $z$ via
 $\phi\sim z^\Delta$ . 

Before we go into specifics, let us first
remind that all  operators are divided into two
classes: (i) The ``protected" ones have zero anomalous dimensions
while (ii) ``unprotected"
ones -- a generic case -- have nonzero ``anomalous" dimensions
resulting from field interaction and depending on the coupling.
This distinction will be very important in what follows.

 Bulk wave eqns for spin-S fields have a generic form 
 \be \partial_z e^{-B}  \partial_z  \phi(z) +(m_4^2-M_5^2/z^2)  e^{-B} \phi(z)=0\ee
where 
 \be  B=z^2+(2S-1)ln(z) \ee
 We follow notations of KKSS \cite{Karch:2006pv}, except that
we have added extra
 5-d mass term.  
  Standard substitution
 $\phi=e^{B/2}\psi$ transform this into a Schreodinger-like eqn without first order
 derivatives
 \be  -\psi'' +V(z)\psi=m_4^2\psi \ee
 with \be V(z)=z^2+2(S-1)+{S^2+M_5^2-1/4 \over z^2}
 \label{eqn-pot} \ee
 KKSS needed only bound state wave functions,
 given in terms of  Laguerre polynomials.   Quadratic IR potential was tuned  by
\cite{Karch:2006pv} to reproduce nice Regge trajectories : for absent bulk mass $M_5=0$
they are linear and  $m_4^2=  4(n+S)$. 

   The absolute units we will use also follow from
    KKSS notations,  can be fixed
 by calculate some physical mass.
 Rho meson is an example of the $protected$ state, associated with
 conserved vector current: it has
  with $M_5=0$ and $S=1$: a solution without
 nodes (n=0) gives $m_\rho^2=4$. Using it as an input we fix our unit of length 
 as \be  length\, unit =2/m_\rho =  0.51 fm \ee.
  The expected position of the  domain wall at large $N_c$ is
 \be z_{dw} \approx 0.4 fm =.777 (length\, unit) \ee 
 Thus
 {\em most of the wave function is in fact located in the strongly coupled
 domain} and modifications due to weakly coupled domain
 at $z<z_{dw}$ turns out to be very small, as far as the 4-d spectroscopy goes.
 However hard processes will be sensitive to this small tail of the wave function,
and their relative normalization is crucial for what follows.
 
Because generic bulk objects are excited
 (closed) string states, there masses 
are non-trivial: in fact their evaluation 
 is one of central subjects  of current string theory.
Interpolating anomalous dimensions (or $M_5$)  between
 weakly and strongly coupled regimes
is a hot subject.
%
%
 Curiously it was not yet discussed in the AdS/QCD
context: papers we know on both holographic spectroscopy and exclusive processes simply
ignore them,  using  canonical (bare) dimensions. There is no conservation laws to protect them most of the operators -- e.g. scalars or baryons -- 
from acquiring anomalous dimensions, and one may ask if 
the usual form of AdS/QCD models can be used in those case.

 So, what is actually known about the ``bulk spectroscopy".
 While in weak coupling
 the magnitude of the
anomalous dimensions is  $O(\lambda)$ small, in strong coupling
 they are in general large. As  argued by Gubser et al, 
  \cite{Gubser:1998bc} generic or ``vibrational"
  string states have anomalous dimensions of the order  
  \be \Delta- \Delta_0\sim (\lambda^{1/4} \ee 
but (to my knowledge) their theory is not yet developed.
There is more progress for ``rotational" string states, especially
for operators with large flavor 
content (large angular momentum $J$ in $S_5$) which developed
into exact spectroscopy using Bethe-ansatz methods.

 Much less is  done for the operators with the usual spin $S$.
 In the case of   large spin $S$
anomalous dimension grow only logarithmically
\be \Delta - S=  f(\lambda)ln(S/\sqrt{\lambda})\ee
and $f(\lambda)$ seems to be universal function for cusps: its large coupling
limit $f(\lambda)\rightarrow {\sqrt{\lambda}/ \pi}$
correspond to folded spinning strings as shown
in \cite{Gubser:2002tv}.   For orientation, if the strongest QCD coupling is 
  $\lambda \sim 80$, and thus this function can reach 
$f(\lambda)\sim 3$ or so: thus  expected  anomalous dimensions
may be of the size of several units or so.

  Although 
we dont yet know appropriate  $M_5$ values for all these operators, 
   its quantitative role is easy to understand from the wave equation.
In the effective potential (\ref{eqn-pot}) $M_5$ simply adds to 
``angular momentum" coefficient of  
 the $1/z^2$ term.  Thus increasing $M_5$  shift
 the wave function a bit more into IR (larger $z$),
 increasing the size of the state and its 4-d masses.
More  radical effect is the $decrease$ of the tail
of the wave function at small $z$, important for hard processes.
 We will return to solutions of this eqn in the two-domain scenario, after we  
briefly review the formalism for hard processes.
 

\subsection{Hard processes in AdS/QCD}

  Polchinski and Strassler pioneered treatment of hard
elastic 
\cite{Polchinski:2001tt} and deep inelastic (DIS) 
\cite{Polchinski:2002jw} scattering in the holographic (strong coupling) setting.
They placed the scattering process into the 5-d bulk,
using  for propagating bulk states string-based scattering amplitudes.
As hadronic
wave functions are depending on  $z$ as  described
above, hard
probes  emit
virtual bulk fields as well.  For example, electromagnetic
formfactors can be thought in terms of bulk current-fields 
vertex 
\be FF(Q) \sim \int dz \psi(z) \partial_m \psi  A_m(z)\ee
where bulk photon $ A_m(z)$ has
imaginary 4-mass $m_4=iQ$
and thus for large $Q$ it is  exponentially
decreasing at large $z$
%
%
 $A_m\sim exp(-Qz)$. Since
 hadrons have  wave functions $\phi\sim z^\Delta$
%
their convolution has a 
maximum at 
\be z_* = 2\Delta/Q  \label{eqn_z*}\ee
It means that physical elastic (or exclusive) processes
proceed via
 hadrons  
in a very compressed state,
 to a size $\sim 1/Q$. (For deep inelastic  
process only some part of a hadron of such size is involved.)  

Brodsky and Teramond \cite{Brodsky:2007vk} obtained the following nice
expression for formfactors
\be F(Q^2) = \Gamma(\tau) { \Gamma (1+Q^2/4) \over\Gamma (\tau+Q^2/4)
}
\ee
where $\tau$ is the  twist$\tau=\Delta-S$ of the operator. 
They were able to describe available  experimental 
data for pion and nucleon formfactors, assuming bare
 twist $\tau=2$ for pions and $\tau= 3$ nucleons . 
While the former
can indeed be produced by a conserved axial current with protected
 dimensions, 
 there is no reason to think the nucleon's conformal dimension is
 protected and remains integer. 
 The absolute values of the pion formfactor does not match
  asymptotic perturbative value: and for many other exclusive
  reactions in which the power laws seem to be working there are
  doubts that those are indeed the regime at asymptotically large $Q$.
  A generic qualitative behavior of all such processes will be 
discussed
  in the next section.
  
\section{Wave functions and hard processes in the two-domain scenario}

 Let us first describe the two-domain wave functions we propose to use.
To find them we will need more 
  general solutions
 \be \psi={C_W \over \sqrt{z}} W((m_4^2/4+1/2, M_5/2, z^2))+\nonumber \\ {C_M\over \sqrt{z}} M((m_4^2/4+1/2, M_5/2, z^2))
\ee
 where  $W,M$ are  Whittaker functions and $C_W,C_M$ some constants.

Let me deviate from a well-trotted path here, and instead of canonical
vector or pseudoscalar mesons have use an example more exotic states with larger operator dimension and $M_5$.    For reasons to become clear in the next section,
we will use objects with bare dimension 3. As we dont know its anomalous dimension,
we will use some arbitrary value 
equal 4 (7 for total dimension at strong coupling).
 From spectroscopic point of view this can be e.g. a
  scalar meson  $(\bar q q)$
  
  As an approximate  solution in spirit of tho-domain scenario, one can  
continuously connect solutions with two  different
bulk masses
$M_5$ in the weak and strong coupling domains .
  The 4-d mass should of course be the same,
derived from continuity of the logarithmic derivative at the domain wall
\be  {d \over dz}log\psi(m_4,M_5^{weak},z=z_{dw}) = \nonumber \\
  {d \over dz} log\psi(m_4,M_5^{strong},z=z_{dw})\ee
 Note that the Whittaker function appropriate for l.h.s.
(small $z $) is M, while for r.h.s. 
 (large $z$) it is $W$.
Of course there should be a continuity of both the wave function and
its derivative, with a jump in the second derivative defined by
the jump in the bulk mass.

\begin{figure}[t]
\includegraphics[angle=0,width=8cm]{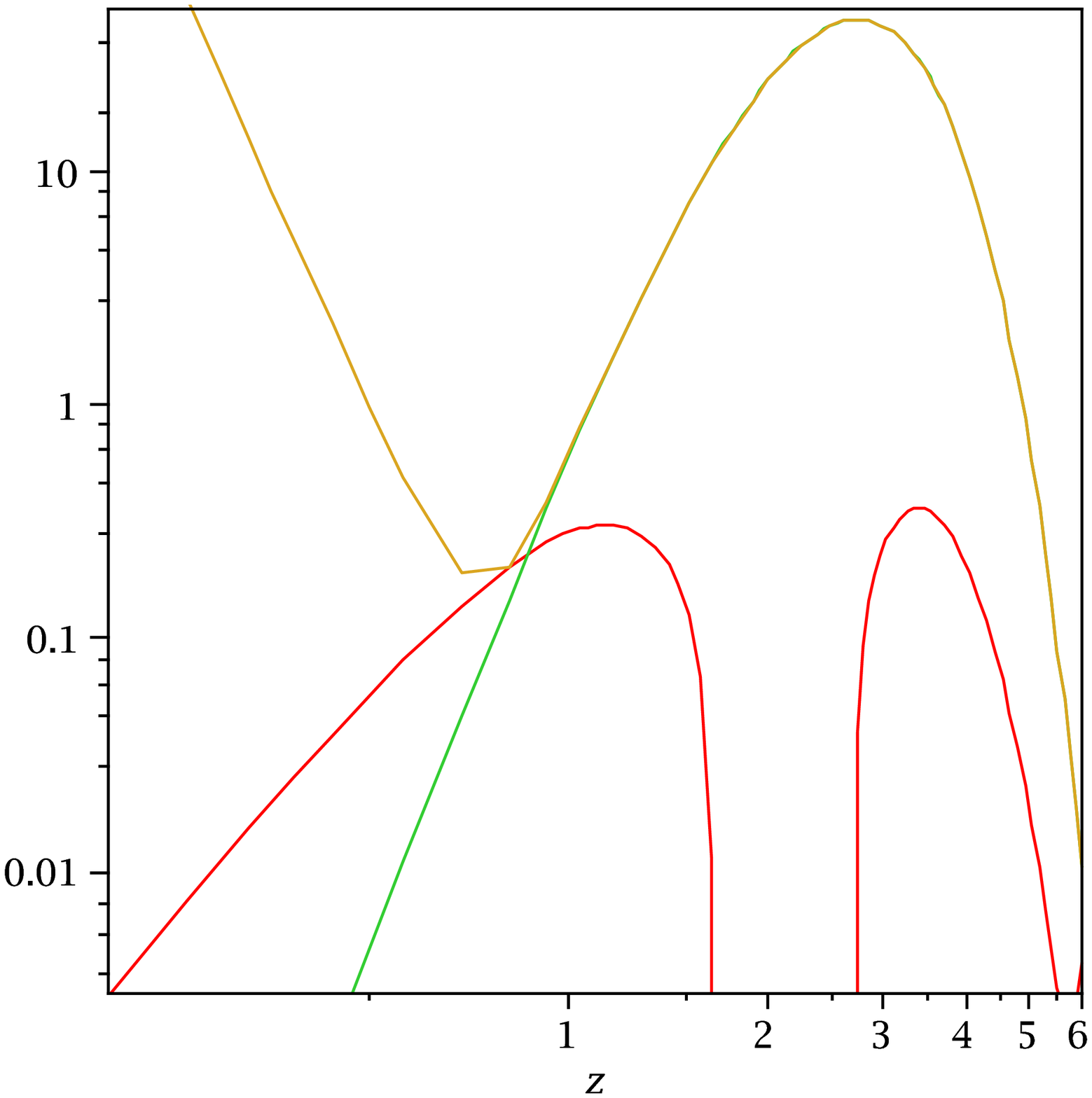}
\includegraphics[angle=0,width=8cm]{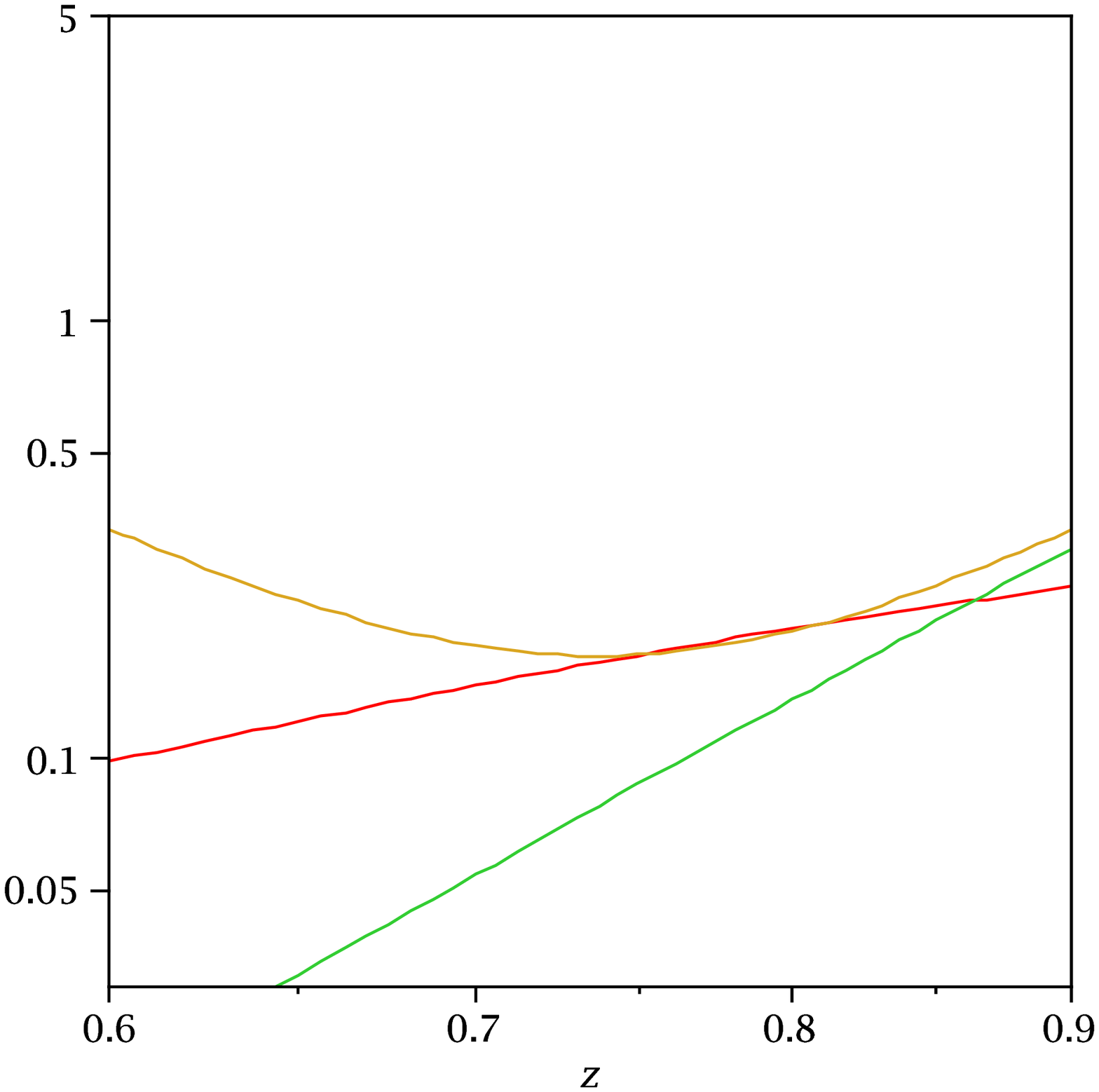}
\caption{(color online) The composite wave function connecting 
soluiton with bare dimension 3
at $z<z_{dw}=.777$ to a solution with dimension 7 at  $z>z_{dw}$.
(a) is the overall picture while (b) is a blow up of the connection region. 
The lowest at small $z$ (green) curve is the dimension 7 solution with
4d mass $m_4^2=14$ which would be there without a weak coupling
domain.  
The next (red)  and the one curving up at small $z$ are dimension 3 and 7
solutions with slightly shifted mass $m_4^2=14-.00005$: they join smoothly
at $z=z_{dw} $.}
\label{fig_wf}
\end{figure}

How it works in practice is shown in Fig.\ref{fig_wf}. Note that in this example
the wave function modified tail is about 4 orders of magnitude down from
its maximum.Although  
the value of the 4-d mass $m_4$ is only changed by a tiny amount, it is enough to influence
the wave function behavior  below $z=z_{dw}$ allowing to
connect two functions smoothly, as shown in Fig.\ref{fig_wf}(b).


The reader may think that tiny tail of the wave function is completely irrelevant.
It is indeed so for the mass and other
bulk parameters of these states, but not for hard (exclusive) processes
which is dominated by small $z$
  (\label{eqn_z*}). As $Q$ grows, it moves from
strongly to weakly
coupled domain, and the behavior of all hard processes changes.
One may ask a question, whether it is possible to even see
this transition and thus locate the ``domain wall'.

The formfactor is calculated from the combined wave function by
\be f(Q)=\int {dz\over z}  \psi^2(z) (QzK_1(Qz))\ee
where the last bracket with Bessel function $K_1$
represent the bulk virtual photon. Schematically, there are three
regions: (i) small $Q$ in which the integral is dominated by all $z$
and is close to the total charge, usually taken as 1, (ii) dominated by
intermediate $z^*>z_{dw}$ in the strongly coupled domain, where
one sees power behavior including anomalous dimension; (iii) 
small  $z^*<z_{dw}$ in which the power law switches to bare operator
dimension.
 
\begin{figure}[t]
\includegraphics[angle=0,width=8cm]{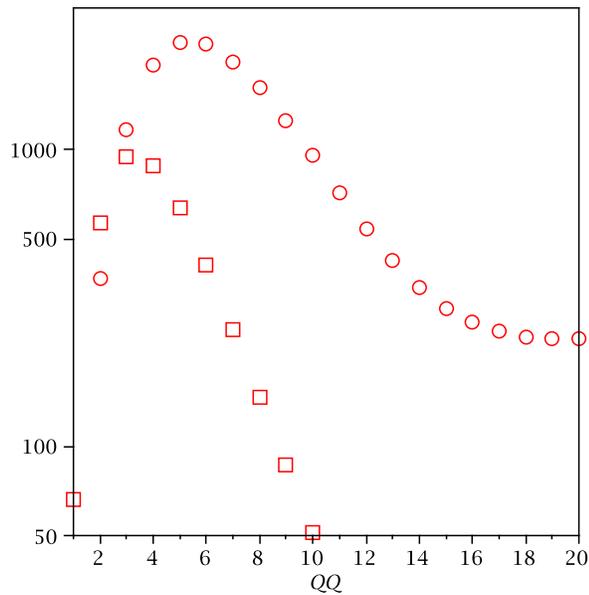}
\caption{(color online) The combination $Q^6 f(Q)$ where
$f(Q)$ is elastic formfactor, versus $Q$ (in units $m_\rho/2$). 
The decreasing curve (boxes)
 is for wave function with the dimension  7.
The curve stabilized at large Q (circles) 
is for the combined wave function, which 
at $z<z_{dw}$ has the (original bare) dimension 3. 
}
\label{fig_ff}
\end{figure}

For the example we considered above, a dimension jumping from
7 in (ii) to 3 in (iii), the formfactor (times $(Q^2)^3$, bare dimension)
is shown in  Fig.\ref{fig_ff}.
It  displays falling behavior (ii) 
 induced by anomalous dimension, which is then
stopped at the r.h.s. of the picture
as formfactor ends up in the 
weakly coupled regime (iii).
 Thus -- at least in this example --
transition from strong to weak coupling is clearly seen in the formfactor.

  (In the next section we will
discuss real-life experimental data for a 
process with bare dimension of the momentum 6,
to which this figure should be compared. It is thus worth noticing that
the value of Q at which the formfactors levels off is as high as 6 GeV.) 

\begin{figure}[t]
\includegraphics[angle=0,width=10cm]{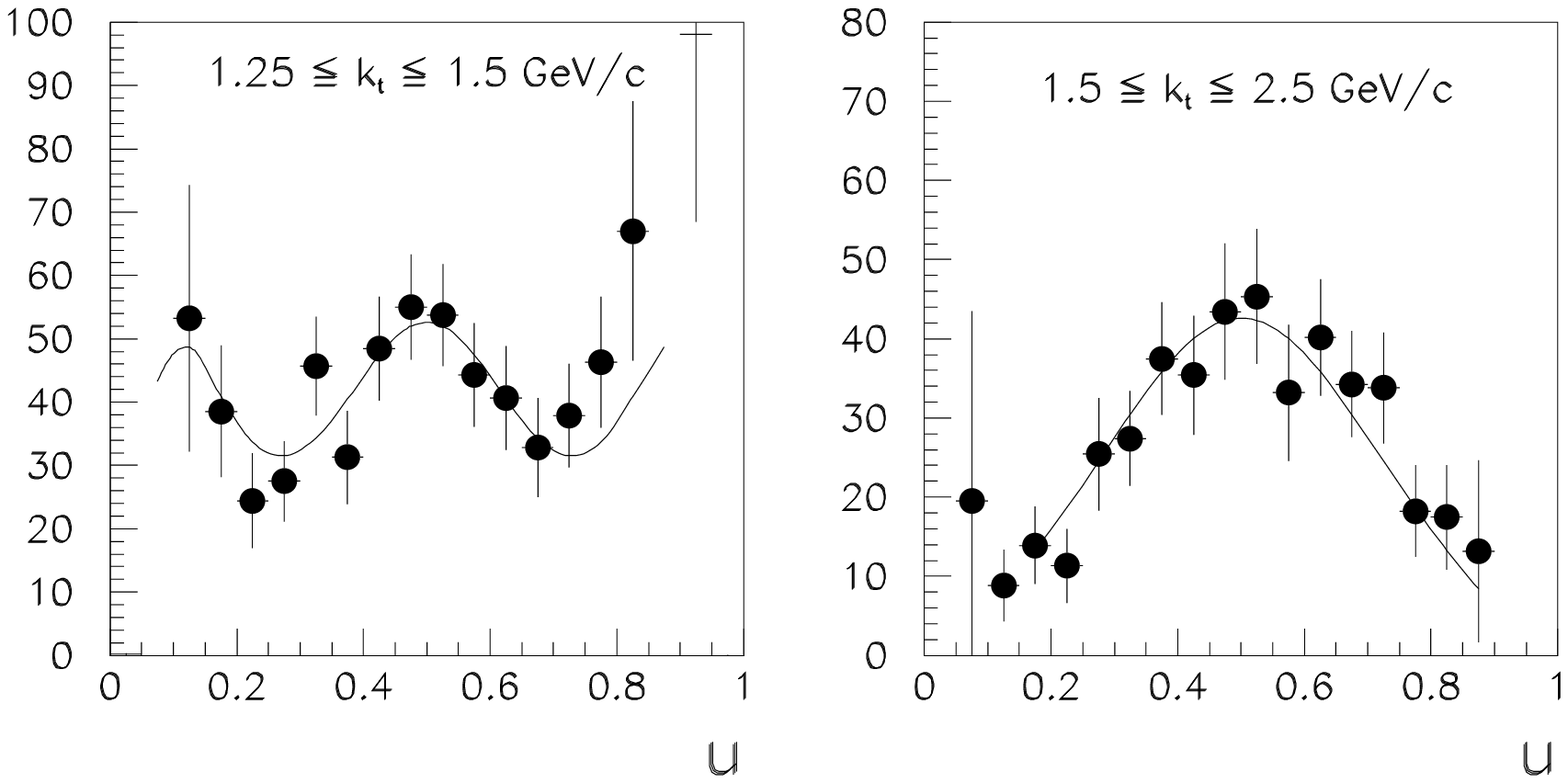}
\vskip -5.cm
\includegraphics[angle=0,width=9cm]{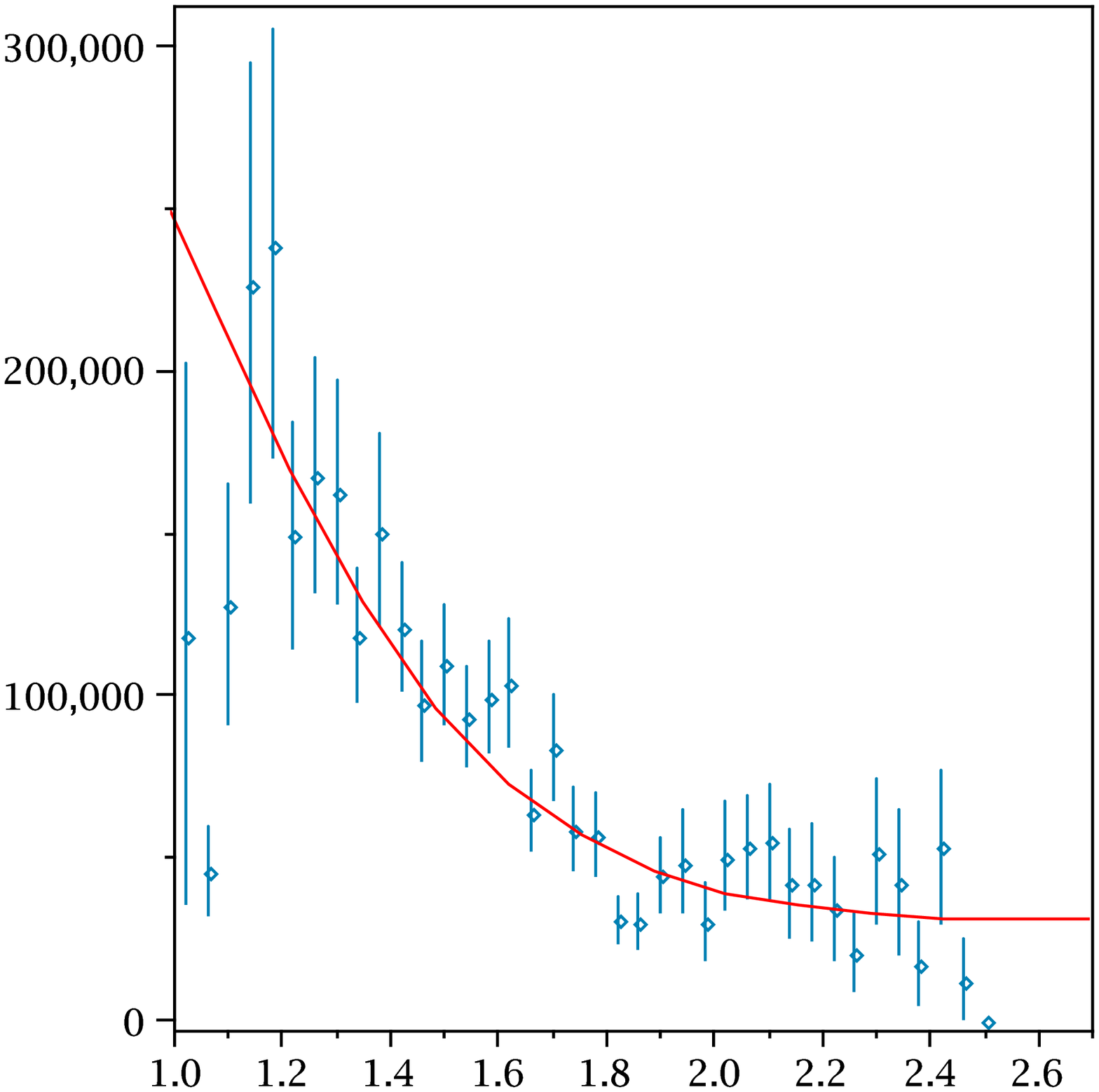}
\caption{(upper) The Acceptance-corrected $u$ distributions of diffractive
di-jets events from  E791 \protect\cite{Ashery}.
 (lower) The points are experimental $k_t$ distribution,
shwn as  $k_t^6 d\sigma/dk_t$
(arbitrary units) vs $k_t$ (GeV). The curve is (rescaled)
formfactor shown in Fig.\ref{fig_ff} by circles: it is plotted to 
test the agreement of their shapes.
 }
\label{gegen}
\end{figure}
\subsection{Is the ``domain wall''   observed already?}
\label{sec_excl}

 How one can observe rapid jump in coupling/anomalous dimensions
 is already discussed
above and displayed in
   Fig.\ref{fig_ff}.  The outlined procedure predicts uniquely
a shape of the transition, from anomalous to bare dimensions
seen in hard processes, as well as the magnitude
of the processes. The transition happens when 
 the peak of the convolution (\ref{eqn_z*}) crosses $z_{dw}$
as $Q$ increases. Ideally it should be
 reflected in the cross section of any hard process,
as all QCD processes
 depend on the coupling.

The simplest object of the kind is the pion formfactor.
Unfortunately experiment has not yet seen how transition
to old perturbative prediction takes place,
as it is very hard to measure it at large $Q$. 
All we know is that the observed $Q^2F_\pi(Q)$  at $Q\sim 1-2\, GeV$
is about twice large than the asymptotic value,
so some decrease and then leveling at new level should happen.
As far as I know, nobody have seen it on the lattice either. The 
 magnitude of the pion formfactor at $Q\sim 1-2\, GeV$ induced by
 instantons has been calculated 
in ref. \cite{FSS}, but  
unfortunately 
the approximation made
are not good at large enough $Q$,
so the transition to the perturbative
regime has not been seen in this approach as well. 
More or less similar situation is with the nucleon formfactor
and many other exclusive reactions.

Another hard reaction involving the pion is
 pion diffractive dissociation into two jets 
\be \pi \rightarrow  jet(k_1)+ jet(k_2)\ee
 first
theoretically discussed by Frankfurt et al \cite{Frankfurt:1993it}
and studied experimentally by
the Fermilab experiment E791. It seems to be showing  
 a  transition from the non-perturbative to the perturbative
regime we are looking for.
 Large transverse momenta
of the jets ensure that a pion is found in a small-size configuration,
with $q$ and $\bar q$ at  small distance
$\sim 1/k_t$ from each other. Due to the so called color
transparency effect, this small color dipole interacts weakly
with the nuclei. In fact in the actual
 experiment the nuclear recoil to the nuclei used
(Pt one of the targets) is very small, less or comparable to
$1/R_{nucleus}$: the process is thus completely coherent
over all nucleons, with the cross section  $\sim A^2$. 
Literally, a pion squeezes through all small holes between
the nucleons of the target, and makes a coherent wave after that.

Fig.\ref{gegen} 
displays two main findings of E791  \cite{Ashery}:
\\ (i) In  Fig.\ref{gegen}(upper),
one sees that the  distribution over a jet momentum fraction $u$ 
 is radically different in two
momentum windows, the  one at higher $k_t$
consistent with perturbative or
``asymptotic'' distribution $u(1-u)$.\\(ii)   Fig.\ref{gegen}(lower) shows that two (somewhat shifted) windows
also have radically different $k_t$ dependence.
 The one at higher $k_t>1.9\, GeV$ is consistent  with
bare dimension of momentum 6 \cite{Frankfurt:1993it}, while
the lower one can be fitted with the  power 10.

 The lower figure should  be compared with
Fig.\ref{fig_ff} above, and the shapes are really very similar.
Again, the interpretation 
  is that the $k_t>1.9 GeV$ 
regime corresponds to perturbative
 asymptotics domain, while the lower $k_t$ are affected by 
 the strongly coupled region where anomalous dimension
is high. The transition is quite clearly seen in the data,
 which we take as the first experimental evidence in
favor of a rapid weak-strong transition. Note that in order to compare it with
formfactor one should use 
$Q=(2k_t)$ which is $\sim 4\, GeV $ at the transition point.

Many open questions to diffractive dissociation include: 
Can the dimension jump
be indeed accounted for by some anomalous
dimension? Does it correspond to expected  jump of the coupling,
from $\alpha_s=1/2$ to 2?
Is it related to instantons? Can ``strongly coupled''
distribution over fraction of the energy $u$ at $k_t<1.5$ window
Fig. \ref{gegen}(left upper)
be also explained? How similar diffractive data would look like for a kaon
or photon beams, or for a proton dissociating into 3 jets?

\subsection{Beyond the two-domain scenario}

 Completing discussion of the wave functions and hard processes, let
us briefly comment on how in principle one should
 get a more consistent solution of the problem. based not
on a pictures with jumps but  appropriate equations
describing evolution of the coupling and anomalous dimensions.

The renormalizability of the theory implies that the equations
depend on $current$ coupling $\lambda(z)$
which itself changes due to first-order 
renormalization group equation 
\be {d\lambda \over dz} = \beta(\lambda(z))\ee
(Relation to
  second order eqn for the dilaton which follows
from the Lagrangian is explained
 in  \cite{Gursoy:2007cb}. )
Similar first order
eqns are expected for
 the bulk (mixing) masses of a set of fields $\phi_A,A=1..K$
with the same quantum numbers should be
 given by a similar evolution
equations depending on the local coupling
\be {d M^2_{AB}(z) \over dz} = \gamma_{AB}(\lambda(z))\ee
Since those are
the first order eqns, one has to specify just their
initial values, which are 
bare canonical 
dimensions at 
the boundary $M_{AB}(z=0)=0$. Combining these two eqns one
gets a generic solution
\be M^2(\lambda) =M^2(0)+\int_0^\lambda d\lambda' {\gamma(\lambda')\over \beta(\lambda')}  \ee

\section{A ``domain wall'' made of instantons}
\label{sec_inst}
\subsection{Size distribution of  QCD instantons}
  In the introduction we have mentioned puzzles related with
the ``chiral scale'' and large non-perturbative
corrections in all spin-zero channels, unaccounted by the OPE.
  These puzzles were resolved  by realization \cite{Shuryak:1981ff}
that small-size  instantons $\rho\sim 1/3 \, fm$   are in fact
very abundant in the QCD vacuum. In spite of
%
 small  coupling making
 the semiclassical exponent frighteningly  small 
\be exp(-8\pi^2/g^2(\rho))\sim exp[-10]  \ee
large preexponent and attractive interactions increase it
substantially, leading to actual instanton diluteness
parameter
\be n_{I+A}\bar \rho^4\sim 10^{-2}\ee
(here $n_{I+A},\bar \rho$ are 
the instanton (plus antiinstanton) density and mean radius,
respectively.

Eventually, the so called Interacting 
Instanton Liquid Model (IILM) was developed,
which included
't Hooft multi-fermion
interaction to all orders. It
 provided quantitative deception of
instanton effects including chiral symmetry breaking
and various correlation function, for review see \cite{Schafer:1996wv}.

\begin{figure}[t]
 \includegraphics[width=8cm]{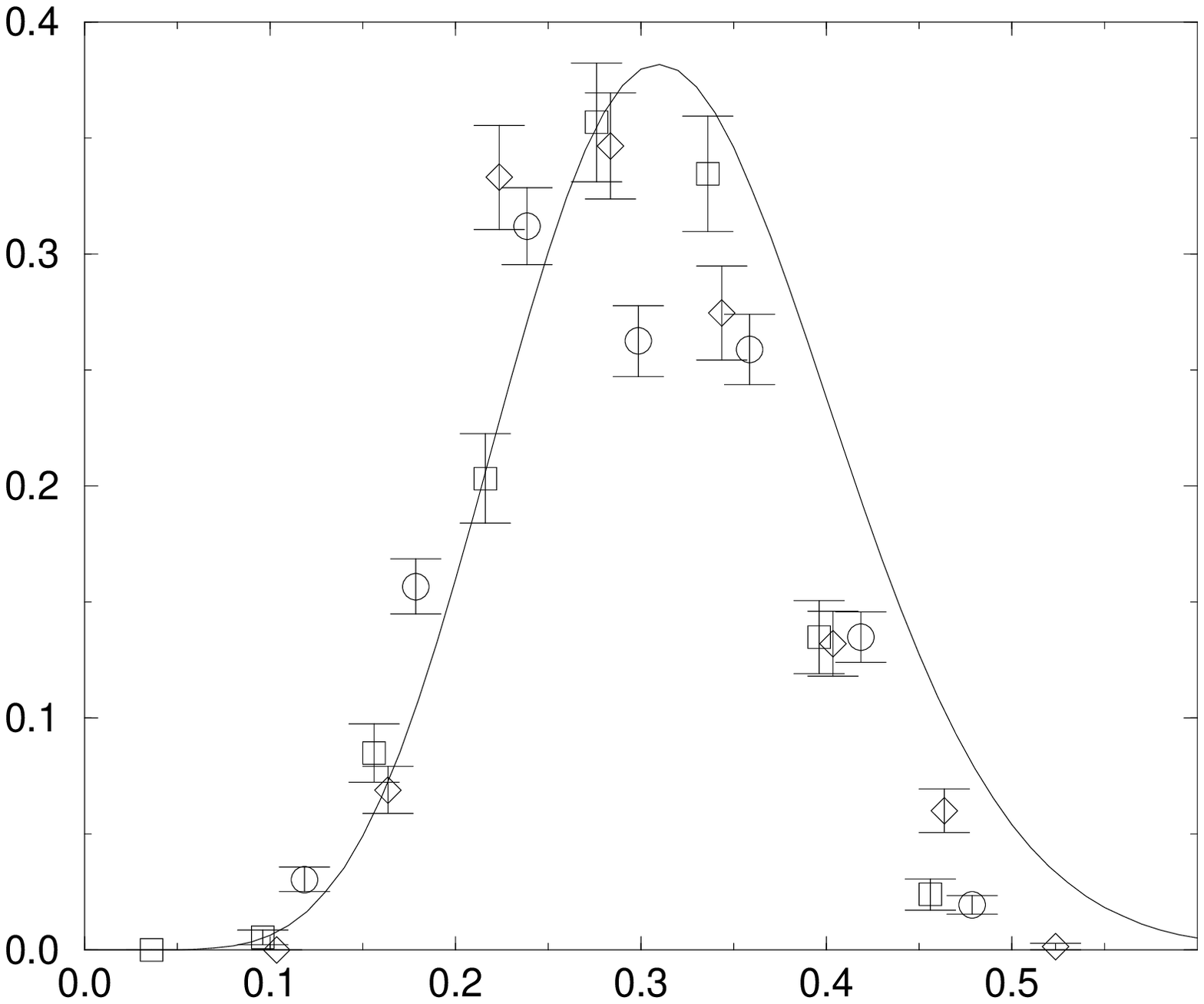}
  \includegraphics[width=8cm]{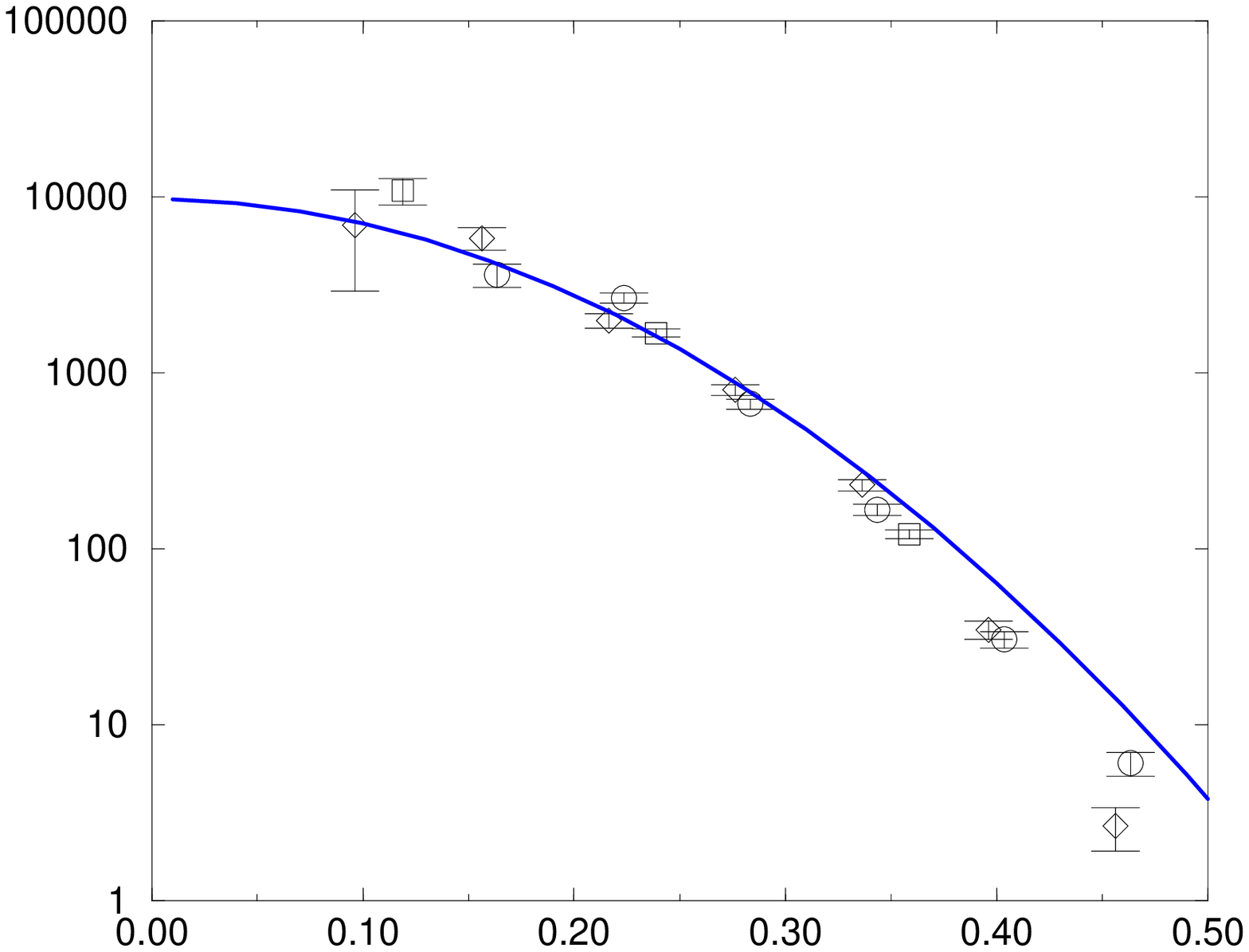}
  \vspace{-.05in}
  \caption{
(a) The instanton density $dn/d\rho d^4x$, [fm$^{-5}$] versus its size
 $\rho$ [fm]. (b) The combination  $\rho^{-6} dn/d\rho d^4z$, in which
 the main one-loop UV behavior drops out for $N=3,N_f=0$, thus
 presumably giving only the IR part of the potential we discuss.
 The points are from the lattice study \protect\cite{anna},
for pure gauge theory, with 
$\beta$=5.85 (diamonds), 6.0 (squares) and 6.1 (circles). (Their
comparison should demonstrate that results are
rather lattice-independent.)
The line corresponds to the 
expression $\sim exp(-2\pi\sigma\rho^2)$, see text.
  } \label{fig_sizes}
\end{figure}

    For the purpose of this paper we will not need to discuss its
phenomenological consequences: we instead
 focus instead on one issue, the distribution over the
instanton size $\rho$. While asymptotic freedom
at  strongly suppresses instantons in UV (small $\rho$), there is no
commonly accepted
explanation of the suppression which is observed  in IR: see e.g.
lattice data  shown in
Fig.\ref{fig_sizes}.
I suggested \cite{Shuryak:1999fe}
 that it is due to dual Higgs appearing
in ``dual superconductor'' picture of confinement.
As can be seen from the lower figure, both the quadratic
dependence on $\rho$ and the coefficient in 
the predicted correction factor
\be \label{eqn_large_rho_suppr}
{dN\over d\rho}/{dN\over d\rho}|_{semiclassical}=exp(-2\pi\sigma \rho^2)\ee
(with $\sigma$ being the QCD string tension) 
agree well with the available lattice
data.

\begin{figure}[t]
\includegraphics[width=5.4cm]{rho_nc_scal.eps}\\
\includegraphics[width=6.4cm]{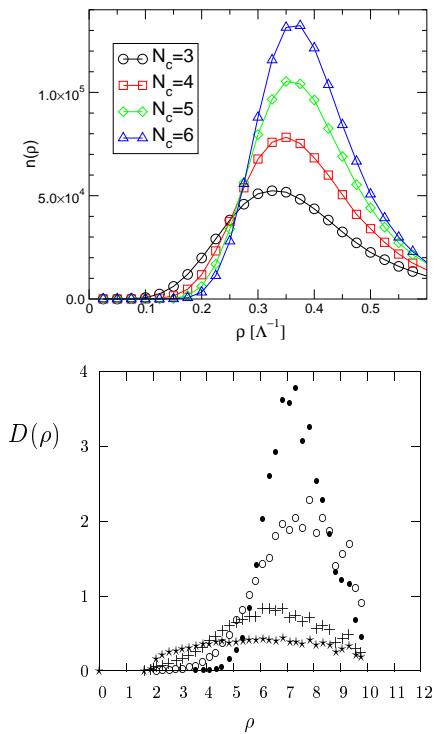}
\caption{\label{fig_size_Nc}
The figure on the left shows the instanton size 
distribution obtained from numerical simulations of the 
instanton ensemble in pure gauge QCD for different numbers
of colors \protect\cite{Schafer_Nc}.
 The figure on the right shows lattice results 
reported by Teper \protect\cite{Teper}. The $\star+\circ\bullet$ 
symbols correspond to $N_c=2,3,4,5$.}
\end{figure}

 The main point we want to make by this point is that
 the
instantons are rather well
localized in $\rho$  in the $N_c=3$ case.
  For larger number of colors $N_c\rightarrow \infty$
 this tendency is further enhanced, as seen from
 Fig.\ref{fig_size_Nc} which compares IILM 
calculation
 by
T.Schafer \cite{Schafer_Nc} and 
lattice from M.Teper \cite{Teper}.

Since small size instantons are suppressed
exponentially in $N_c$,
 $\exp(-8\pi^2/g^2)\sim (r\Lambda_{QCD})^{(11/3)N_c}$, the left side
of these peaks are getting more and more steep at larger $Nc$.
There is clearly strong suppression from the 
right (IR) side as well, but its nature and $N_c$ dependence
remains unclear. The ``dual superconductor'' strength
per color and thus the string tension
remain finite at  large  $N_c$, and
eqn (\ref{eqn_large_rho_suppr}) predicts fixed suppression on
the IR side. The Instanton liquid calculation in upper figure
do not have a dual superconductor VEV but mutual repulsion of
instantons, which work similarly: this a ``triangular''
shape of the curves at large $N_c$. What is the shape of the lattice
data (lower curve) is hard to tell: pragmatically it looks like
 a delta-function.

\subsection{Instantons in weakly coupled \N=4 theory}
\label{sec_inst_ads}
   Instantons in
weakly coupled \N=4 theory  
were extensively studied by
Dorey,Hollowood,Khoze and Mattis, see review 
 \cite{Dorey:2002ik}. 
 The absence of running coupling
decouples the tunneling amplitude from the size dependence.
They have shown that
 in this theory instantons naturally live in $AdS_5*S_5$
space. The size $\rho$ is simply identified with the 5-th coordinate 
$z$, and the size
distribution is given simply by
$dN/d\rho\sim d\rho/\rho^5$ which is nothing but invariant
volume element given by the $AdS_5$ metric. 

This fits perfectly into string language in which instantons are
point-like objects in 5d known as $D_{-1}$ branes. They freely float
in the bulk, filling it homogeneously like dust filling a room.
Furthermore, on a quantum level one gets perfect fits as well.
Instanton's ``hologramm'' on the boundary, readily
calculated from the dilaton/axion
bulk-to-brane propagators, gives  the correct image
$G^2_{\mu\nu}(x)\sim \rho^4/(x^2+\rho^2)^4$.
A point object cannot be coupled to graviton:  thus
there is no holographic stress tensor. And indeed, instantons
 have it zero $T_{\mu\nu}=0.$ because of self-duality.

 Furthermore, the
remaining 8 supersymmetries rotating the instanton solution
nicely relate  fermionic zero modes to bosonic ones.
The remaining $S_5$ 
space  also nicely come out of  mesonic variables associated with
fermionic zero modes.

\begin{figure}[t]
\includegraphics[width=8cm]{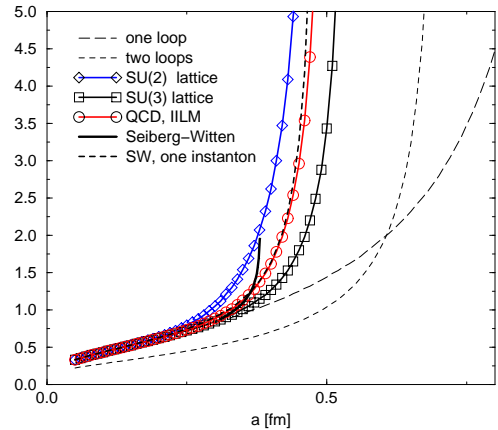}
\caption{
 The effective charge $b \,g^2_{eff}(\mu)/8\pi^2$ (b is the coefficient
of the one-loop beta function) versus normalization scale $\mu$ (in units of
its value at which the one-loop charge blows up). The thick solid line
correspond to exact Seiberg-Witten solution  for the \N=2 SYM, the thick dashed line
shows the one-instanton correction. Lines with symbols (as indicated on figure)
stand for N=0 QCD-like theories,
SU(2) and SU(3) pure gauge ones and QCD itself. Thin long-dashed and short-dashed lines are one and tho-loop results.
}  \label{fig_coupling}
\end{figure}

\subsection{Instanton-induced corrections to
coupling, in QCD and SUSY theories }
So far we treated instantons as some point probes, 
 while now  we would ask
{\em What is the back reaction of instantons on the running coupling?}.
 This is in fact a very old question: 
Callan,Dashen and Gross  30 years ago derived its
first order in instanton density, and even 
attempted to resum its nonlinearly.
A decade ago the issue was looked at by Randall, Rattazzi and myself
\cite{Randall:1998ra}, from which we borrowed Fig.\ref{fig_coupling}.

 The external field is supposed to be normalized at some normalization scale
$\mu$, and following Callan-Dashen-Gross idea
we included all instantons with size $\rho
<1/\mu$. The effective charge is then defined as:
\begin{equation} \label{eq_CDG}
{8\pi^2 \over g^2_{eff}(\mu)} = b \, \ln({ \mu\over \Lambda_{pert}}) 
 \nonumber \\
- {4\pi^2 \over (N_c^2-1)} \int^{1/\mu}_0 dn(\rho) \rho^4 
({8\pi^2 \over g^2_{eff}(\rho)})^2 
\end{equation}
  where 
$b=11N_c/3-2N_f/3$ is the usual one-loop coefficient of the beta function, and $dn(\rho)$ is the distribution of instantons 
(and anti-instantons) over  size.

In it we compare QCD results (based on size distribution from
the lattice already discussed) to both first order instanton
correction and the
exact results, for the
\N=2 (Seiberg-Witten) theory.
This is seen from the exact result for the effective coupling 
\begin{equation}
{8\pi \over g^2(u)} = {K(\sqrt{1-k^2}) \over K(k)}
\end{equation}
where K is the elliptic function and its argument has
\begin{equation}
k^2={(u-\sqrt{u^2-4 \Lambda^4})\over (u + \sqrt{u^2-4 \Lambda^4})}
\end{equation}
being the function of
the gauge invariant vacuum expectation value of
the squared scalar field 
\begin{equation}
u={1\over 2}<\phi^2>={a^2 \over 2} +{\Lambda^4 \over a^2} + \ldots
\end{equation}
where $a$ is just its VEV.
 For large $a$ there is a weak coupling expansion
which includes  instanton effects 
\begin{equation} \label{pert}
{8\pi \over g^2(u)}={2 \over \pi} \left ( \log \left( {2 a^2 \over \Lambda^2 }\right) -
 {3 \Lambda^4 \over a^4}+ \ldots \right)
\end{equation}

The behavior is shown in  Fig.(\ref{fig_coupling}), where we
have included both a curve which shows the full 
coupling (thick solid line), as well as a curve which illustrates
only the one-instanton correction (thick dashed one). Because we will want
to compare the running of the coupling in different theories,
we have plotted $b g^2/8 \pi^2$ (b=4 in this case is the one-loop coefficient
of the beta function) and measure all quantities 
in  units of $\Lambda$, so that  the one-loop
charge blows out at 1. 
Note the very rapid change of the coupling induced by instantons.
It is also of interest that the full multi-instanton sum makes
the rise in the coupling even more radical than with only the
one-instanton correction incorporated.

 One more argument in favor of rapid instanton-driven 
transition comes from analytic results on the
\N=4 theory on a different 4-dim manifold called $K_3$
by Witten and Vafa \cite{Vafa:1994tf}, who
were able to sum up all 
instanton contributions. Their results show that
there is a rapid transition 
from a ``dilute'' to ``dense'' instanton phase, at the critical
coupling
$g^2/4\pi=1$. Furthermore, it becomes 
a true phase transition at large $N_c$ \cite{Papadodimas}.  
This fact further suggests
 that the phenomenon is generic,  presumably taking
place
for all gauge theories, supersymmetric or not.

\subsection{Instantons in AdS and correlators   }
 After this extended introduction, let us discuss
 how instantons can be incorporated into the
AdS/QCD framework. 

 Two first steps are clear: (i) We would  like to keep
metric to be that of pure $AdS_5$ and (ii) identify
the instanton with a bulk point, with its $z$ coordinate identified
with the instanton size $\rho$.

 Of course instanton distribution over $z$ in AdS/QCD 
should  be completely different from that in AdS/CFT:
 instanton ensemble is no longer a kind of a dust,
filling space homogeneously, but is instead
 located in a narrow region of $z$ 
 shown in Figures above.
 This is of course
 caused by the running coupling, or in the AdS/QCD language,
 by the z-dependent
dilaton potential. In IR it should be quadratic,
 in harmony with KKSS, and in UV it should match the
perturbative beta function. 
In the large $N_c$ limit the instanton size distribution
 gets infinitely sharp.

  Thus we come to the central idea of this paper:
{\em  the ``domain wall'' is perhaps made of instantons}.


 Let us test this idea, starting with 
its simplest applications.  Consider for example 
the instanton contributions
to gluonic pseudoscalar (scalar)
correlation functions
\be \Pi(x,y) =<O(x)O(y)>\ee
where $O_{PS}={\alpha_s \over 8\pi} G \tilde G$ and  $O_{S}$
has $GG$ without dual. 

 The ``old-fashioned'' way to calculate it
is simply to include the semiclassical fields of the
instantons in the correlator, and integrate over instantons.
  The answer is instructive to write 
 for the Fourier transform
of this (Euclidean) correlator
\be \Pi_{S,PS}(Q)=\pm \int {d\rho d(\rho)\over \rho^5} [Q^2 \rho^2 K_2(\rho Q)]^2
\label{eqn_GGcorr} \ee
where $d(\rho)$ is the instanton size distribution
and the bracket is Fourier distribution of instanton's $GG(x)$.

\begin{figure}[t]
 \includegraphics[width=8cm]{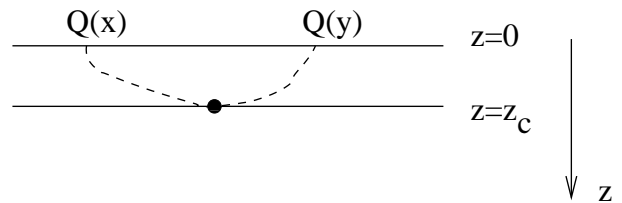}
  \caption{The diagram showing instanton contribution
to the topological charge correlator. The dashed lines
are axion (or dilaton) propagators, and the
black circle is the instanton.  At large $N_c$  it must be
located at the domain wall $z=z_{dw}$.
  } \label{fig_GGcorr}
\end{figure}

  This same result appears in the AdS/QCD model framework,
where it
 comes from the diagram  shown in Fig.\ref{fig_GGcorr} and
the square bracket in (\ref{eqn_GGcorr}) is reinterpreted 
as a bulk-to-boundary propagator
of the axion/dilaton. 
It was 
discussed by Katz and Schwartz \cite{Katz:2007tf} in 
 the AdS/QCD model, which
 instead of the instanton density
$d(\rho=z)$  have some  ``anomaly coupling'' $\kappa$.
These authors rightly noted that confinement-related
 IR cutoff of the AdS space
automatically solves the divergence over large instanton sizes:
 but they did not mentioned that if their $\kappa$
be constant in their Lagrangian this very integral would  
power diverges at $small$ $z$! Indeed, their version 
of AdS/QCD  is conformal in UV and it does not know
about the asymptotic freedom.

In our scenario ( in the large $N_c$ limit)
all instantons are at $z=z_{dw}$,
and thus no power divergence in the correlators.
In fact the delta-function distribution is exactly
 what I used for the simplest 
 ``random instanton liquid model'' 
 \cite{Shuryak:1981ff}  in early 1980's to understand correlators
  of this kind. 
   Two parameters
of the model were found to be the (4d) density $n_I=n_A\approx .5 fm^{-4}$  
and the size (now the domain wall position) $z_{dw}=1/3\, fm$.
They have  quantitatively described a lot of phenomenology, including the $\eta'$
mass and mixing with glueball channels, which were 
among the puzzles
pointed out in \cite{Novikov:1981xj}.

What about other correlation functions? As we already mentioned,
vector and axial currents have been
well described by Schafer \cite{Schaefer:2007qy} in AdS/QCD
models without instantons. Those are two exceptional cases,
protected operators which indeed long known to ignore
instanton scale $1/3 fm$. In our two-domain scenario
it means that corresponding bulk fields do not care for the domain
wall
as they are 
 protected and have no anomalous dimensions on both
sides of it. 

Generic operators however should see the domain wall. An indeed,
as was first documented in detail in review \cite{Shuryak:1993kg},
all scalar/pseudoscalar operators split precisely at the instanton
scale  $z_{dw}=1/3\, fm$. The 
  AdS/QCD models which are conformal in UV --
such as discussed
by  Schafer \cite{Schaefer:2007qy} or Katz and Schwartz
  \cite{Katz:2007tf}) --  dont even know
about  this scale and would not agree with the QCD phenomenology.

\subsection{Domain wall made of interacting instantons}

The partition function of ILM can be be  written as
\be
\label{Z}
 Z = \frac{1}{N_+!N_-!}\prod_i^{N_++N_-}C(N_c)^{N_++N_-}\int {d\Omega_i d^4x_i
 dz_i \over z_i^5} \nonumber \\
  exp\left[-S_{eff}(z_i) -S_{int}^{glue} -S^{f}_{weak}  -S^{f}_{strong}
\right]\ee
where we have identified summation over instantons and antiinstantons
with proper integration over collective coordinates
mentioned and orientations in
color space $\Omega_i$. 

 The action includes single-body effective action $S_{eff}(z_i)$
which includes classical $8\pi^2/g^2$ and quantum corrections,
all depending on $z$ because of running coupling (in UV) and
confinement (in IR).
The interaction we split into three parts, the gluonic one and
two fermionic parts, which we explain subsequently.
The bosonic action $S_{int}^{glue}$ for a pair of two instantons
(or antiinstantons)
is
the log of the moduli space metric (the overlaps of bosonic
zero modes). For instanton-antiinstanton pair there are no 
relative zero modes and interaction
 is determined from the so called ``streamline''
equation: the details can be found in a review \cite{Schafer:1996ws}.
The only important point is that
 collective coordinates
appear in two   combinations
\be \label{zd}
cos(\theta_{IA})=(1/N)Tr[\Omega_I
 \Omega^+_{A} (\hat R_\mu\tau_\mu)],\,\,\,
d_{IA}^2={(x_I-x_A)^2\over z_I z_A} \ee
The former is called the relative orientation factor, 
the orientation matrices
$\Omega$ live in $SU(N)$ while the $\tau_\mu$ is the usual
4-vector constructed from Pauli matrices: 
 it is nonzero only in the 2*2 corner of the $N*N$ color space.
The 
$\hat R$ is the unit vector in the direction
of 4d inter-particle distance $x_I-x_A$. The second
 combination of position and sizes appears
 due  to conformal invariance of the classical YM theory.
  
\begin{figure}[t]
 \includegraphics[width=8cm]{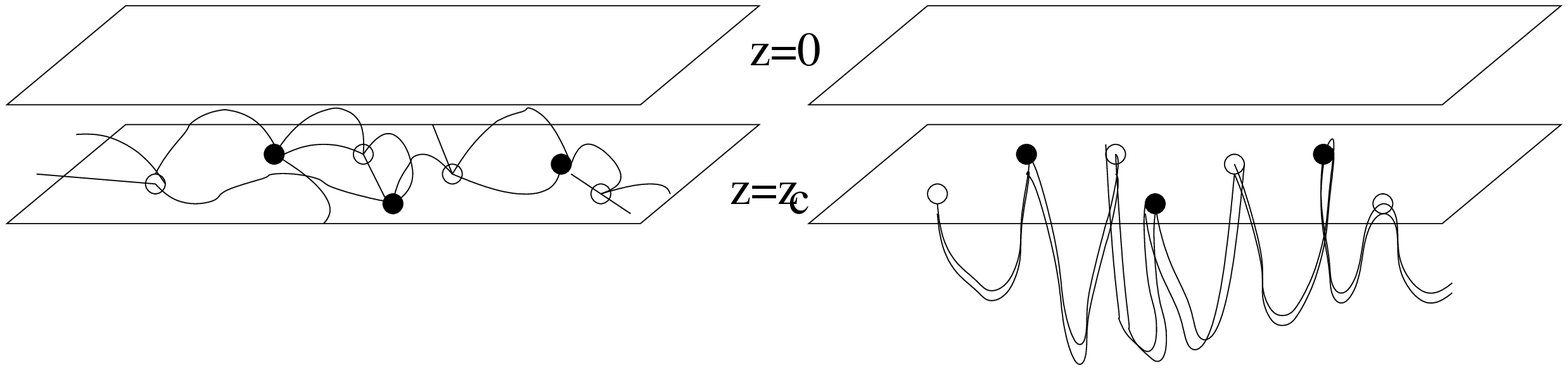}
  \caption{Schematic  structure of the ``instanton wall'':
black and open points indicate point-like instantons and
antiinstantons. The left and right pictures show
interaction between them, in weakly and strongly coupled
domains.   We assume two massless quark flavors, so the lines
on the left -- 4 per instanton --
 are exchanged zero mode quarks. Double lines in the right
figure are mesons.
 } \label{fig_wall_inside}
\end{figure}

The most complicated (nonlocal) part of the effective action
is due to fermionic exchanges, originated from 
the determinants of the Dirac operator
\be  exp\left[ -S^{f}_{weak}  -S^{f}_{strong}
\right]= \Pi_f det(\bar \rho (iD\!\!\!\!/+im_f))\ee
In the spirit of two-domain description, we have split it into
two parts, which keep track of two different regimes.
Their pictorial  explanation
in  Fig.\ref{fig_wall_inside} is also for clarity  split
into two parts.

The crucial part of this determinant includes a subspace
spanned by the fermionic zero modes: thus instantons
can be viewed as effective point-like vertices
with $2N_f$ legs, known as 't Hooft vertex.

In the {\em weakly coupled domain}
 quarks propagate independently and the
picture corresponds to Fig.\ref{fig_wall_inside} (left).
One can approximate
quark propagation between instantons by free fermion propagators
and right
the corresponding part of the Dirac operator as
\be 
\label{D_zmz}
exp(-S^{f}_{weak})=\Pi_f det \left (
\begin{array}{cc}  im_f & T_{IA}\\
                   T_{AI} & im_f 
\end{array} \right ).
\ee
with $T_{IA}$  defined by
\be
\label{def_TIA}
T_{IA} &=& \int d^4 x\, \psi_{0,I}^\dagger (x-x_I)iD\!\!\!\!/
\psi_{0,A}(x-x_A),
\ee
where $\psi_{0,I}$ is the fermionic zero mode.
The terms proportional to quark masses would be
mentioned only
 in section \ref{sec_open}: they have a similar overlap matrix element
but without $D\!\!\!\!/$ and for zero modes of II or AA types.
Note that each matrix element has  the meaning of a ``hopping
 amplitude'' of a quark
from one pseudoparticle to another. 
Furthermore, the determinant of the matrix  (\ref{D_zmz}) is also equal
to a sum of all closed loop diagrams, and thus ILM
 sums all orders in 't Hooft effective $2N_f$-fermion effective
interaction. 
 The matrix element can be written as 
a function of the same two variables (\ref{zd}) as the gauge action
\be
\label{TIA_sum_par}
  \bar \rho T_{IA} = cos(\theta_{IA}) f(d_{IA})\hspace{1cm}
f\approx
  \frac{4 d_{IA}}{\left(2+d_{IA}^2\right)^2}.
\ee
where the expression is an approximate simple parameterization
of a more complicated exact result. 

In the strongly coupled domain quarks propagate in form of
(colorless) mesons, with
picture shown in Fig.\ref{fig_wall_inside} (right).
The 'tHooft vertex can be rewritten in the Nambu-Jona-Lasinio
form, which for two massless flavors is \cite{Schafer:1996wv}
\be  S^{f}_{strong} \sim   ({2N_c-1 \over 2N_c})
[(\psi^+ \tau_a^- \psi)^2+(\psi^+ \gamma_5\tau_a^- \psi)^2] 
\nonumber \\
+({1\over N_c})
 (\psi^+ \sigma_{\mu\nu}\tau_a^- \psi)^2\ee
where 4-d flavor matrix $\tau_a^-=(\tau_1,\tau_2,\tau_3,-i)$,
including e.g. in the pseudoscalar term both the pion term and -- with
the opposite sign due to $U(1)_A$ breaking - the $\eta$.
It only includes gamma matrices
$1,\gamma_5,\sigma_{\mu\nu}$ which correspond to chirality flip:
thus no vector/axial mesons. Ignoring subleading in $N_c$ last term
one may use only 4 spin-zero mesons, $\pi,\eta,a_0,f_0$.

This bosonized 'tHooft vertex is now quadratic for two flavors:
thus the structure of lines in Fig.\ref{fig_wall_inside} (right).
The fact that this form should be used {\em only below the domain
  wall}
$z>z_m$
corresponds to the old UV cutoff of the NJL model
at $\Lambda_\chi$. The reason we show
mesonic lines propagating rather deep into large-$z$ IR region
is based on the structure of the scalar propagator
in the $AdS_5$ between two points with the same $z$
\be D\sim {z^4 \over (z^2+x^2)^4}\ee
which decreases with the 4-d distance $x$ much stronger than 
with the 5-th coordinate $z$.
That is presumably the reason for the fact (mentioned in the
introduction)
that one cannot cut off $z$ till say 10 fm or so, without
some observable consequences for QCD observables.


 Summarizing this section: one can  naturally rewrite
instanton ensemble in the $AdS_5$ language.
The classical measure $d^5x/z^{-5}=d^5x\sqrt{-g}$ is the
right invariant measure in $AdS_5$.
 The variable $d_{IA}$ on which 
all the functions involved depend can 
be interpreted as
the
 invariant distance between 2 points in 
the   AdS$_5$ metric. So one can view all complicated interactions
 of the point-like bulk objects in $AdS_5$, {\em without}
referring back to the boundary.

\section{Discussion}

\subsection{Mesons as vibrations of the wall }

Top-down approaches  introduce ``matter branes''
on which quarks live, starting with the $D_7$ branes by
Karch and E.~Katz\cite{Karch:2002sh} and then $D_8$ branes
%
of the
Sakai-Sugimoto model\cite{Sakai:2004cn}. In the latter the
right and left parts of massless quarks
on two separate
sets of $N_f$ branes, but chiral symmetry breaking
unite them into a single one, with some ``connecting part''
related right and left-handed components together. 
The position of this part fixes the ``constituent quark'' mass
scale.

Our ``domain wall'' is a dynamically 
formed structure, but it is destined to play the same role
as the ``connecting part'' of the matter brane. 
Indeed, the ``zero mode zone'' quarks who lives near it
  are responsible for
 the chiral condensate and get constituent quark mass.
 Furthermore, as experience
with instanton liquid calculations shows,
the quarks belonging to  ``zero mode zone''  dominantly
contribute to  the lowest hadrons,
from pseudoscalar and vector mesons to even the nucleon.
Lattice studies such as \cite{Negele,DeGrand} confirmed that
keeping only Dirac eigenstates with the lowest eigenvalues
(and made of instanton zero modes) in the propagators
is sufficient to reproduce hadronic correlation functions 
at large distances associated with those hadrons.

And yet brane constructions have provided a
 completely new view at the   lowest pseudoscalar and vector mesons,
identifying them with
 collective vibrations of the brane as a whole. 
This valuable idea can be applied to the domain wall as well.
 The
effective action for mesons in brane constructions
 follow from the so called DBI action, 
 which is some tension times the invariant
area of the brane calculated in a background metric.
 A vibrating ``domain wall''
 also  generate mesons,
and the corresponding 
effective action can include the area term as well. However
 it also should depend on the dilaton potential
(which provides the minimum  responsible
  for its fixed position $ z_{dw}$ 

\be S_{eff}\sim \int d^4x d\tau 
e^{-V_{wall}(z-A_z(x,\tau))}
 \sqrt{-det(g_{wall})}\ee

The fact that  ``domain wall'' is not a brane but
 has a finite width at finite $N_c$, and even in
 the 
 infinite $N_c$ it is not a continuous object but
just a 4d cloud of correlated point objects
is not a problem: low lying oscillations with 
large wavelength are just {\em collective vibrations
of a wall as a whole}.
 ( AdS/QCD  should not 
have true brane-like objects formed, as 
there is no  supersymmetry to hold it together.)

In the leading quadratic order the effective action can only be
a hidden-gauge-like action $S\sim \int d^4x(F_{\mu\nu})^2$,
 but the nonlinear terms should not be the same as there would be
different contributions from the dilaton potential.


\subsection{Other open issues }
\label{seq_open}

{\bf What is the geometry of other extra dimensions?} 
The \N=4 theory has them neatly put into a sphere $S_5$,
which also nicely comes from mesons related to fermionic zero modes
\cite{Dorey:2002ik}. One certainly do not expect in QCD 
any relation between flavor and these extra dimensions.
There is however an important relation between instanton
topology and extra dimensions, which is revealed
by color ``Higgsing'' of the \N=4
theory, namely starting with $N_c$ original $D_3$
branes put not into the same  point but rather into
certain  $N_c$ points on a circle $S^1$. 
Thus color group is broken and monopoles can be introduced 
as open strings with ends on different branes. Instanton topology
correspond to  circles of  $N_c$ segments, going around  $S^1$.  
Although the radius of the circle can be later considered vanishing,
this interpretation of the instantons 
as (small string) circles in extra dimensions 
is worth keeping.


{\bf What is the role of relative color orientation vectors? Can
one get rid of them in a more generic formulation?}
 Both bosonic and fermionic determinants in QCD
partition function above contain some remnants of the color -- the
 relative color orientation vectors.  Pragmatically, those can
be viewed  as some kind of classical group-valued
``spin variables'' associated with 
otherwise point-like instantons in the $AdS_5$ bulk.

 Perhaps a more natural
 formulation (which does not exist yet) is to combine
the circle topology just mentioned with
Kraan-van Baal picture of (finite-T) instantons as
being made out of $N_c$ dyons. Not only can  a classical solution
be understood like that, but,
 the gauge part
of the instanton measure itself  can be exactly rewritten
in terms of $4N_c$ coordinates of $N_c$ dyons:
see \cite{Diakonov:2005qa} in gauge theory and 
 \cite{Lee:1997vp} in brane language.
In the monopole language the
 ugly color angles are gone. One remaining task
--perhaps not too hard -- is
is to do so for all fermionic factors in the measure,
for cases other than \N=4 theory. More formidable task is
to do monopole manybody dynamics, which is orders of magnitude more
difficult than the statistical mechanics of the ``instanton liquid''.

In summary, one can rewrite rules for QCD instanton ensemble in
the $AdS_5$ setting, although it does not look nearly as nice
as for the \N=4 theory. Another much larger task may be to
include instanton sector as a part of some dual formulation
altogether emphasizing magnetically charged objects
--monopoles and dyons.

{\bf What is the meaning of $z$-dependent axion field? }

 Gursoy, Kiritsis and Nitti \cite{Gursoy:2007er,Gursoy:2007cb}
have included in their version of AdS/QCD
an axion field $a(z)$, which (like the dilaton) have some $z$-dependence
due to  potential induced by interaction with other bulk fields.
However the physical meaning of 
 ``the running $\theta$ parameter'', as they call it, is not obvious
 and deserves to be discussed.
 
Standard rules of holography require that the UV boundary value
 $a(0)$ be identified with input UV value of the $\theta$ parameter,
and the coefficient of  $z^4$ with the topological charge density
operator $G_{\mu\nu} \tilde G_{\mu\nu}$. 

Although the QCD vacuum is known to have near-zero value of the $\theta$ parameter,
discussion of ensembles with the nonzero topological charge / $\theta$
were done theoretically and numerically (on the lattice).
Furthermore, Zahed and myself \cite{ZS_sphalerons} argued that 
the fireballs created in heavy ion collisions
at RHIC should have $Q\sim 10$ because of the order
of a 100 vacuum instantons/antiinstantons  are excited into sphalerons
and some random fluctuations in $Q$ are expected to happen.
We did not know how one can observe it: but 
 recently Kharzeev
 and Zhitnitsky \cite{Kharzeev:2007tn}
proposed charge asymmetry driven by electric field along
the fireball angular momentum at nonzero $\theta$, which can do it. 
There are even 
preliminary indications from STAR experiment that this effect may in fact
be already observed.

The nontrivial
requirement put forward by GKN is that they  
must be related to each other, via a new condition  \be a(z_{IR})=0\ee
at the infrared singularity.  
It is not clear why this is the case.
Generally the fact that the (topological) charge density and $\theta$
($exp(i\theta)$ is the instanton fugacity) 
is expected. The relation should follow in a standard way from the
free energy (axion action, or vacuum energy) 
\be <Q>=\partial S/\partial (i\theta) \ee
In particular, for noninteracting instantons one gets the usual
vacuum energy $\sim cos(\theta)$ and thus $ <Q>\sim sin(\theta) $,
with zero charge at both $\theta=0,\pi$. 

The axion is a bulk dual to $\eta'$ in 4d. Naturally in the large $Nc$
limit it is massless both in 4d and and in the AdS bulk.
It interacts with instantons and antiinstantons with plus/minus way,
and resumming all axion interactions should look like
4d ``scalar Coulomb gas'' advocated by  Zhitnitsky for some time \cite{Zhitnitsky:1999ki}.
If Nc is not large then $\eta'$ meson has nonzero
(and in fact large)  mass: this corresponds to correlated instantons-antiinstantons
and short screening length for Q. 

Now, any consideration of
 a state with a $nonzero$ average topological charge 
(or $\theta$-induced effects)  must refer to the case when
all  quark masses are nonzero. If so, the 'tHooft vertices shown 
 in Fig.\ref{fig_wall_inside} should be complimented by those
not shown there, with
masses and chirality flips 
(described by the overlap integrals without the Dirac operator, $\sim
 m_f/x^2$). As a result, quark
determinants remain nonzero even for different
number of instantons and antiinstantons.

As a result, the difference
in the total density between instantons and antiinstantons
leads to a different 
their interaction as well. 
  If also $N_c$ is finite, 
 instantons and antiinstantons will not have the same
distribution in 
sizes (their z-coordinates) and thus z-dependence of the
 axion field naturally appears.
Thus in general the GKN's introduction of  $a(z)$ does 
make sense: but 
only if $all$ quark masses are nonzero.
How this important condition appears in their approach
is unclear.

\section{Summary}

 We argued that a good approximate geometry for AdS/QCD
may be two  $AdS_5$ domains, a weakly  and a strongly coupled ones,
separated by the ``domain wall''. We  identify its position
with old ``chiral scale'' (\ref{eqn_chiral}). We
further propose a dynamical
explanation for the ``domain wall'', relating it to mean instanton
 size.
We further argued that  the domain wall should become sharp  in the large
$N_c$ limit, in which all instantons
get localized at the same value  $z=z_{dw}$. 

The gravity-dual description is only useful in the
strongly coupled domain, while the usual weak-coupling
 perturbative/semiclassical methods should be used in
in the other. As we discussed, bulk properties such as
bulk masses may jump across the domain wall.
But important principle
connecting them is  {\em continuity of all bulk fields}  
required 
across the wall.
Indeed, the wave equation with a jump in the bulk mass
require only a jump in its second derivative.

 Existence of one common normalized wave function of $z$
 is very important for
 hard processes. Their asymptotics may have perturbative (bare)
powers of momentum transfer, but its normalization
 would be still
defined mostly by the wave functions in the strongly
coupled domain. 

The shape of the hard amplitude's dependence on $Q$ is determined
essentially by one number -- the strong-coupling
anomalous dimension/5-d mass.
The fact that this nontrivial shape (circles in Fig.\ref{fig_ff}) is seen experimentally
(lower Fig.\ref{gegen}), if not accidental, suggests two important lessons:
(i) the anomalous dimensions may be rather large, 3-4 units; (ii) and
they can indeed be about constant in the scale interval under consideration.
These are the best hints we have, that something like {\em two distinct
conformal regimes} may in fact coexist in QCD.

 A lot of work would be needed to clarify coupling behavior and the
 role of instantons in large-$N_c$ QCD and in the  structure of the ``domain wall".
 Treatment of the lowest
mesonic modes as collective vibrations of the ``domain wall'' is another
exciting direction to study.

\section*{Acknowledgment}
This paper was initiated 
at the Newton Institute (Cambridge) program "Strong Fields,
Integrability and Strings", Aug-Sept.07, so I thank its organizers
 and especially Elias Kiritsis,Misha Stephanov and Arkady
Vainshtein
for discussions.
It is also pleasure to thank a long-time friend Stanley Brodsky, who 
helped to guide me through literature, and Dan Ashery
for providing the  E791 data.

This work is  supported by the US-DOE grants DE-FG02-88ER40388
and DE-FG03-97ER4014.

\end{document}